\documentclass[fleqn,usenatbib]{mnras}

\usepackage[T1]{fontenc}

\DeclareRobustCommand{\VAN}[3]{#2}
\let\VANthebibliography\thebibliography
\def\thebibliography{\DeclareRobustCommand{\VAN}[3]{##3}\VANthebibliography}

\usepackage{graphicx}	
\usepackage{amsmath}	
\usepackage{amssymb}	

\usepackage{newtxtext, newtxmath}

\title[Unveiling the outer dust disc of TW Hya]{Unveiling the outer dust disc of TW Hya with deep ALMA observations}

\author[J. D. Ilee et al.]{John~D.~Ilee$^{1}$\thanks{E-mail: J.D.Ilee@leeds.ac.uk (JDI)},
Catherine~Walsh$^{1}$,
Jeff~Jennings$^{2}$,
Richard~A.~Booth$^{3}$,
Giovanni~P.~Rosotti$^{4}$,
\newauthor
Richard Teague$^{5}$,
Takashi Tsukagoshi$^{6}$ and
Hideko Nomura$^{7}$
\vspace{0.3em}
\\
$^{1}$School of Physics and Astronomy, University of Leeds, Leeds LS2 9JT, UK\\
$^{2}$Institute of Astronomy, University of Cambridge, Madingley Road, Cambridge, CB3 0HA, UK\\
$^{3}$Astrophysics Group, Imperial College London, Blackett Laboratory, Prince Consort Road, London SW7 2AZ, United Kingdom\\
$^{4}$School of Physics \& Astronomy, University of Leicester, University Road, Leicester, LE1 7RH, UK\\
$^{5}$Center for Astrophysics | Harvard \& Smithsonian, 60 Garden St., Cambridge, MA 02138, USA\\
$^{6}$Division of Radio Astronomy, National Astronomical Observatory of Japan, Osawa 2-21-1, Mitaka, Tokyo 181-8588, Japan\\
$^{7}$Department of Earth and Planetary Sciences, Tokyo Institute of Technology, 2-12-1 Ookayama, Meguro, Tokyo, 152-8551, Japan
\vspace{-0.25cm}
}

\date{Accepted 2022 April 29. Received 2022 April 26; in original form 2021 December 17}

\pubyear{2022}

\begin{document}
\label{firstpage}
\pagerange{\pageref{firstpage}--\pageref{lastpage}}
\maketitle

\begin{abstract}
The radial extent of millimetre dust in protoplanetary discs is often far smaller than that of their gas, mostly due to processes such as dust growth and radial drift.  However, it has been suggested that current millimetre continuum observations of discs do not trace their full extent due to limited sensitivity.  In this Letter, we present deep (19\,$\mu$Jy\,beam$^{-1}$) moderate resolution (0.37\arcsec) ALMA observations at 1\,mm of the nearest protoplanetary disc, TW Hya. Using the visibility analysis tool \texttt{frank}, we reveal a structured millimetre intensity distribution out to 100\,au, well beyond previous estimates of 60--70\,au. Our analysis suggests the presence of a new millimetre continuum gap at 82\,au, coincident with similar features seen in optical/near-infrared scattered light and millimetre molecular line observations.  Examination of the fit residuals confirms the presence of the previously reported au-scale continuum excess at 52\,au ($\mathrm{P.A.}=242.5\degr$).  Our results demonstrate the utility of combining deep, moderate resolution observations with super-resolution analysis techniques to probe the faintest regions of protoplanetary discs.

\end{abstract}

\begin{keywords}
protoplanetary discs -- stars: individual: TW Hya -- techniques: interferometric -- submillimetre: planetary systems
\vspace{-0.3cm}
\end{keywords}



\section{Introduction}

Observations of planet-forming discs have shown that the radial extent of their millimetre dust component is, generally, smaller than the extent of molecular gas in the disc \citep[e.g.][]{Panic2009, Andrews2012}.  Initially, this difference was attributed to limited sensitivity \citep[see][]{Dutrey1998}, but further observations revealed that the gas and dust cannot be reproduced with the same surface density profile \citep{Hughes2008}.  In particular, observations with the Atacama Large Millimetre/submillimetre Array (ALMA) demonstrated that a sharp decrease in dust surface at the outer edge is required \citep[e.g.][]{deGregorio-Monsalvo2013, Andrews2016}.

A proposed explanation for these differing radial extents involves the growth and inward drift of dust grains through the disc.  Observations of discs at cm wavelengths have confirmed that dust can grow to mm-cm sizes \citep[see, e.g.,][]{Rodmann2006,Ricci2012} and that the maximum grain size increases toward the inner disc \citep{Perez2012, Tazzari2016}.  These characteristics are predicted by models of growth and radial drift \citep{Birnstiel2010,Birnstiel2012}. Indeed, such models naturally result in a sharp outer edge of the dust distribution (or sharp decrease in the dust-to-gas mass ratio) that is compatible with observations \citep{Birnstiel2014}.  However, observed continuum fluxes and radial extents suggest that drift needs to be halted (or significantly slowed) in the majority of discs \citep{Pinilla2012}.  This can be achieved via local pressure maxima induced by substructure in discs, e.g.\ rings and gaps \citep[see][]{Andrews2020}.     

Despite this apparently self-consistent picture, \citet{Rosotti2019_radii} suggested that even the most recent observations of the millimetre dust continuum in discs are still limited by sensitivity.  In contrast to \citet{Birnstiel2014}, they argue that currently measured disc radii do not trace the sharp outer edge in the dust distribution, but rather the radius out to which the grains are large enough to possess significant (sub-)millimetre opacity.  The dust surface density extends further, but in these regions the grains are small and the opacity significantly lower (factor $\sim$10). If this interpretation is correct, then observations of the outer regions of discs in the (sub-)millimetre should reveal faint emission, given sufficient sensitivity.  
\begin{figure*}
    \centering
    \includegraphics[width=\textwidth]{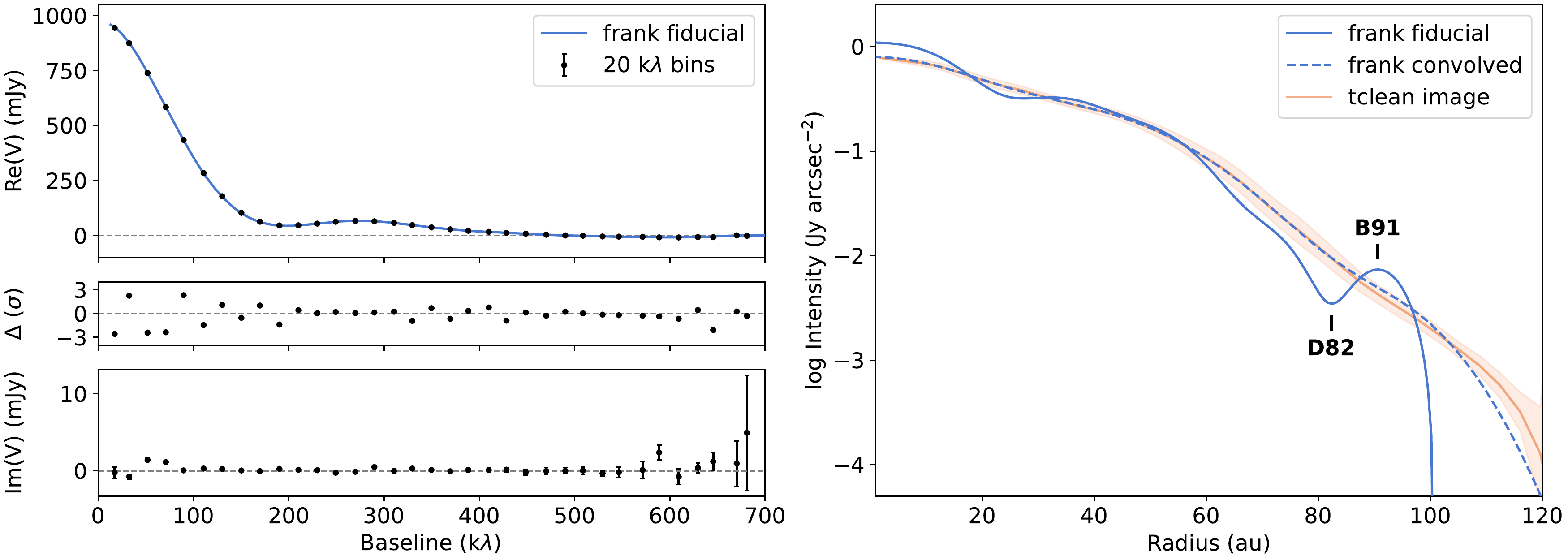}
    \caption{\textit{Left}: Our fiducial \texttt{frank} fit compared to the real component of the visibilities Re(V) shown in 20~k$\lambda$ bins.  The lower panels show the residuals between fit and data ($\Delta$) in terms of the uncertainty on the weighted mean $\sigma$, and the imaginary component Im(V).  \textit{Right:} The radial intensity profile derived from our fiducial \texttt{frank} fit (solid blue) shown alongside the intensity profiles of the deprojected \texttt{tclean} image (orange) and the fiducial frank fit convolved with the \texttt{tclean} beam (dashed blue).  Shaded regions give the standard deviation for each bin.
    The newly identified dark gap (D82) and bright ring (B91) are marked.}
    \label{fig:frank_fit}
\end{figure*}

Such observations require optimal targets for which the surface brightness sensitivity can be maximised.  TW Hya is a young star of spectral type M0.5, stellar mass $\sim$0.6\,M$_{\odot}$ and age $\sim$8\,Myr \citep{Sokal2018}.  It is surrounded by a well studied protoplanetary disc, due to its close proximity (59.5~pc, \citealt{Bailer-Jones2018}) and near face-on inclination (5\degr, \citealt{Huang2018}).  The TW Hya disc has been characterised across multiple wavelengths, from scattered light \citep{Debes2017, vanBoekel2017} to both thermal dust and molecular line emission \citep{Andrews2016, Teague2017, Huang2018, Tsukagoshi2019, Nomura2021, Macias2021}.  Each of these investigations has revealed a diverse array of substructure in both the gas and dust within the disc.

In this Letter, we present deep observations of the TW Hya disc at 1\,mm obtained with ALMA (19 $\mu$Jy~beam$^{-1}$, 0.37\arcsec).  We perform our analysis in the visibility plane to increase the effective signal-to-noise (S/N), and characterise the dust continuum emission and substructure in the outer ($\gtrsim$60\,au) disc for the first time.  
\section{Observations}
\label{sec:obs}

TW Hya was observed by ALMA in Band 7 on the 3rd of December 2016 for an on-source time of 5.7 hours in configuration C40-4 under project code 2016.1.00464.S (P.I.\ C.~Walsh). Baselines ranged from 15--704~m with 39--46 antennas depending on execution, with precipitable water vapour measurements of 1.0\,mm.  Quasar J1037$-$2934 was used as both a phase and flux calibrator, while J1058$+$0133 was used as a bandpass calibrator. The correlator was configured for a continuum spectral window from 292.8--294.8~GHz with a spectral resolution of 32.0~km~s$^{-1}$.  Data (self-)calibration and imaging were performed with CASA v5.6.2 \citep{McMullin2007}.  A continuum visibility measurement set was created by combining all channels after flagging those with line emission.  Two rounds of phase self-calibration and one round of amplitude self-calibration were undertaken, improving the peak S/N by a factor of 14.  The phase centre was set to $11^{\rm h}01^{\rm m}51\fs811$,  $-34\degr42\arcmin17\farcs267$ throughout.  Initial continuum imaging was performed with \texttt{tclean} using Briggs weighting (robust = 0.5) and multi-scale synthesis (nterms = 1) with scales of 0, 0.36, 1.8, 4.2 and 6.0 arcsec, respectively.  The synthesised beam was $0.39 \times 0.36$ arcsec ($23 \times 21$~au) with a position angle of $-50.1$\degr.  The final continuum rms was 18.8\,$\mu$Jy~beam$^{-1}$ (0.11\,mJy~arcsec$^{-2}$) measured from a signal-free region of the image. We highlight that, in terms of a surface brightness sensitivity per standardised area, our rms is a factor of 160--210 times lower than recent high resolution observations of TW~Hya at similar frequencies (e.g.\ 18.3\,mJy~arcsec$^{-2}$ in \citealt{Huang2018} and 22.8\,mJy~arcsec$^{-2}$ in \citealt{Macias2021}).

\begin{figure*}
    \centering
    \includegraphics[width=\textwidth]{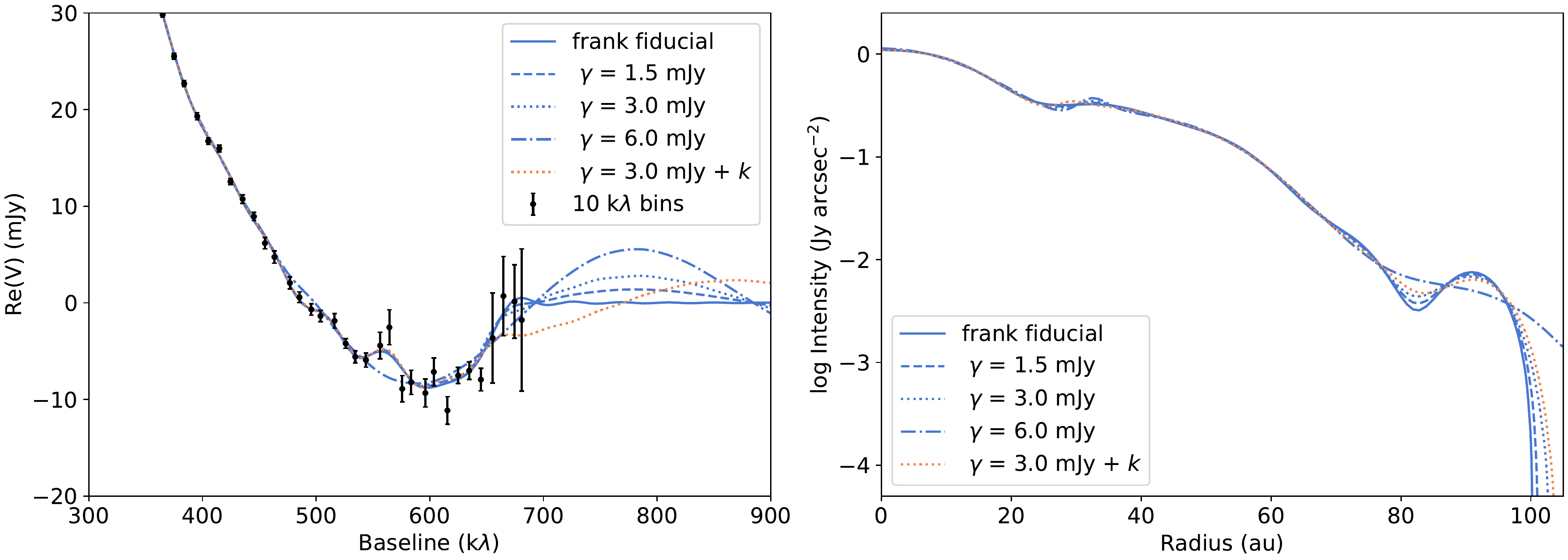}
    \caption{\textit{Left}: Zoom-in of the real component of the visibilities at the longest baselines in our observations, compared with our fiducial \texttt{frank} fit (solid blue) and extrapolations to longer baselines as discussed in Section \ref{sec:robustness}. \textit{Right}: The corresponding intensity profiles based on these extrapolations. }
    \label{fig:uncerts}
\end{figure*}

\section{Analysis}
\label{sec:analysis}

Despite the depth of our observations, the limited spatial resolution of our data (0.37\arcsec, 22\,au) prevents a detailed investigation in the image plane. We therefore fit the visibilities directly with the \texttt{frankenstein} code (\citealt{Jennings2020}, v1.1.0, hereafter \texttt{frank}).  Such analysis can achieve `super' resolution (factors of $\sim$3 better than traditional imaging techniques, \citealt{Jennings2022}) and also acts to improve S/N for axisymmetric structure \citep[see also][]{Walsh2014,Zhang2015,Ilee2020}.

Briefly, \texttt{frank} reconstructs the radial intensity at a set of locations, $r_i$, by using a Discrete Hankel Transform to relate the intensity at these positions, $I(r_i)$, to the visibilities. The intensity is then inferred directly from the observed visibilities using a non-parametric Gaussian Process prior to regularise the fit. The \texttt{frank} code accounts for the inclination of the source by assuming the emission comes from an optically thick and geometrically thin disk, for which the inclination ($i$) and position angle ($\rm{P.A.}$) can be taken account via
\begin{align}
    u = & \frac{u'}{\cos i} \cos({\rm P.A}.) + v' \sin({\rm P.A.}),\\
    v = & - \frac{u'}{\cos i} \sin({\rm P.A}.) + v' \cos({\rm P.A.}),\\
    V = &\, V' {\cos i},
\end{align}
where $V$ is the observed visibility, $u$ and $v$ the corresponding $uv$-points, and primed quantities refer to the de-projected coordinates. We assume the inclination and position angle that have been already well determined from high-resolution observations by \citet{Huang2018}, namely $i=5\degr$ and ${\rm P.A.} = 152\degr$, respectively.  

Figure \ref{fig:frank_fit} (left) shows the comparison between the visibilities and our fiducial \texttt{frank} fit, where we adopt the fit parameters $\alpha=1.2$, $w_{\rm smooth} = 10^{-3}$, $r_{\rm out}=5.0$\arcsec\ and $p_0 = 10^{-15}$\,Jy$^2$ using 300 radial points.  The fit shows good agreement with maximum residuals of less than 3$\sigma$ across all baselines.  Figure \ref{fig:frank_fit} (right) shows the derived intensity profile from this model, along with the intensity profile of the deprojected \texttt{tclean} image and the \texttt{frank} intensity profile convolved to the same beam, demonstrating agreement at all radii.

The \texttt{frank} intensity profile exhibits two features not seen in previous millimetre observations of TW Hya.  Firstly, the emission extends to $\sim$100\,au, well beyond the typical radial extent of $\sim$60-70\,au reported previously \citep[e.g.][]{Andrews2016, Huang2018, Macias2021}. Secondly, we detect new substructure in this outer region of the disc, notably a dark gap centred at 82\,au (hereafter D82) and a bright ring at 91\,au (hereafter B91). We note that the extremely sharp drop of intensity at 100\,au is a result of the log-scaling of the plot and the intensity profile approaching zero.  

\subsection{Robustness of the derived intensity profile}
\label{sec:robustness}

Determining a robust estimate of the uncertainty on the radial intensity profile reconstructed from the visibilities is challenging because intensity reconstruction is formally an ill-posed problem, with the intensity depending on the visibilities at unobserved $uv$-spacings. It is therefore necessary to regularise the fit, which \texttt{frank} achieves via an informative prior that damps power on unobserved small scales. This means that an uncertainty estimated directly from the posterior (here $\sim$0.1\rm mJy\,arcsec$^{-2}$) is a lower bound. Due to the high S/N of the data at the largest observed baselines, the uncertainty due to missing baselines dominates over this estimate.  The high S/N also prevents us from investigating the uncertainty of the fit by varying the hyperparameters $\alpha$ and $w_{\rm smooth}$ as it is insensitive to these choices.

We therefore explore how the \texttt{frank} profile changes when estimating the contribution from the missing long baseline data via several sensible extrapolations beyond the longest observed baselines in our data.  In Figure \ref{fig:uncerts} (left), we show the impact of decaying sinusoidal oscillations with amplitudes of $\gamma=1.5, 3$ and 6\,mJy and a wavelength that approximates the variation in the visibilities between 500--700\,k$\lambda$ (also shifted by half a wavelength, $+k$). We also damp the oscillations on very long baselines by a Gaussian with width 1\,M$\lambda$ to prevent small-scale artefacts in the intensity profile. 

Figure \ref{fig:uncerts} (right) shows the corresponding intensity profiles for these extrapolations, demonstrating how the properties of the disc substructure change.  Our \texttt{frank} fits to extrapolations with smaller amplitudes (e.g.\ fiducial, $\gamma=1.5$\,mJy and $3.0$\,mJy) all recover gap and ring structures with slightly different locations and depths, but a similar goodness-of-fit (calculated across the 400--700\,k$\lambda$ range, which are those sensitive to linear size scales of 30\,au and less).  The gap and ring structure is mostly removed for a \texttt{frank} fit to the $\gamma=6.0$\,mJy extrapolation, but this is a mildly worse reproduction of the visibilities compared to the fiducial fit (where $\Delta \chi^{2} = 12$, corresponding to $2.2\sigma$).  Therefore, while the extrapolations demonstrate that the precise morphology of the substructure depends on the behaviour of the visibilities beyond the largest baselines, we have been unable to remove the substructure entirely and remain consistent with the observed visibilities.  In the absence of a formal uncertainty, we propagate a representative uncertainty based on the $\gamma=3.0$mJy extrapolation in all subsequent analysis.

We also considered other extrapolations, such as a constant visibility amplitude (a point source) and varying the oscillation wavelength by a factor 2. Although not shown, they produce very similar levels of variation as discussed above.  In addition, we tested \texttt{frank} intensity profiles truncated at 80\,au, which produced residual images with $\gtrsim10\sigma$ artefacts out to $\sim$100\,au, further demonstrating the increased radial extent of the disc in our observations.

\section{Discussion}
\label{sec:discussion}

\subsection{The intensity profile compared to previous observations} 

\begin{figure}
\centering
\includegraphics[width=\columnwidth]{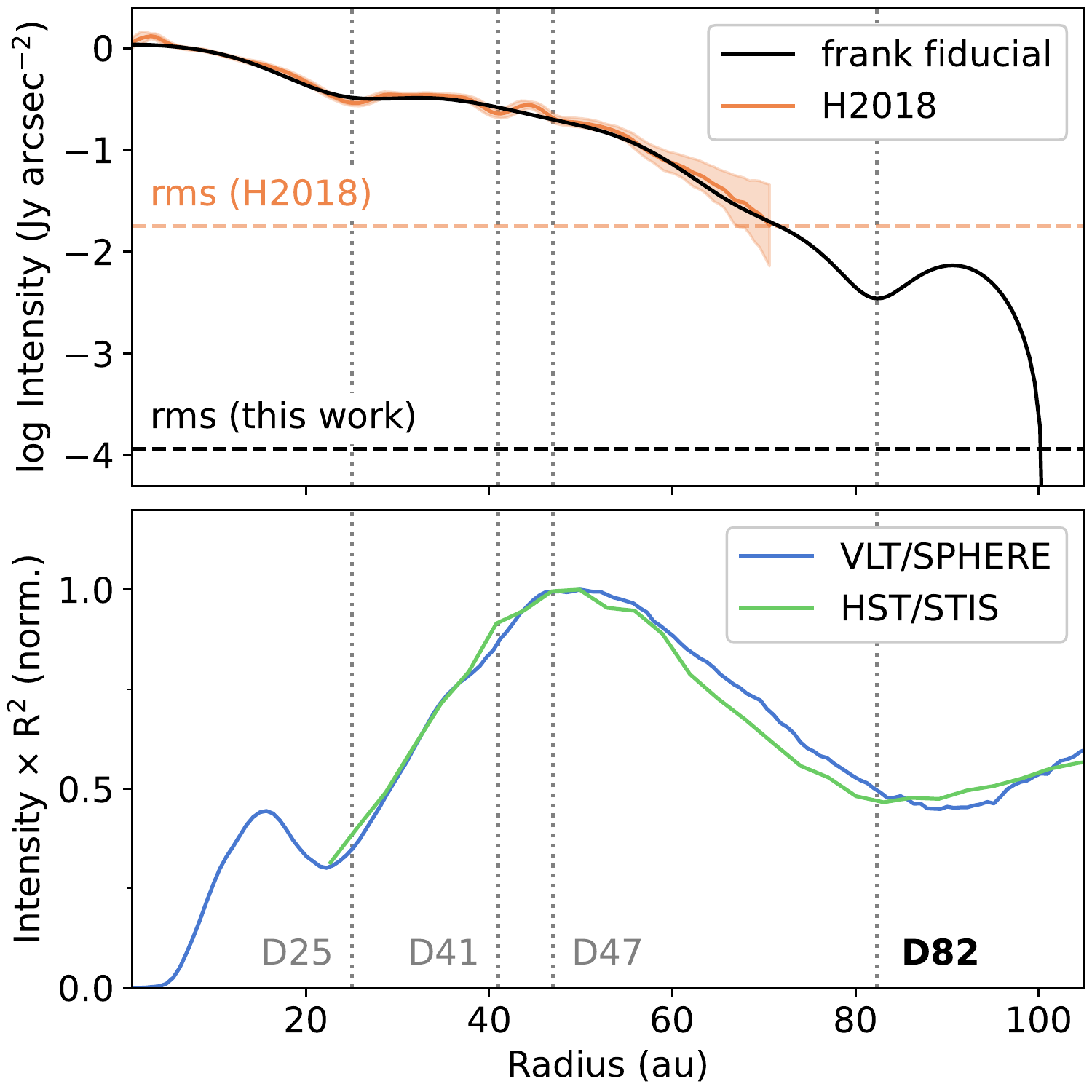}
\caption{\textit{Top}: Radial profile from our fiducial \texttt{frank} fit compared with that from the 290 GHz continuum image from \citet{Huang2018} (H2018).  Image plane rms values are shown with horizontal dashed lines. 
\textit{Bottom}: Radial profiles from scattered light observations with VLT/SPHERE (H-band) and \emph{HST}/STIS (R-band) as presented in \citet{vanBoekel2017} and \citet{Debes2017}, respectively.
Vertical dotted lines mark the location of the previously-identified millimetre continuum gaps (D). The D82 feature coincides with the outer gap seen in scattered light observations.}
\label{fig:radial_profiles}
\end{figure}

\citet{Huang2018} present high resolution (0.13\arcsec) observations of TW Hya at 290\,GHz, providing a useful comparison.  They recover the previously identified continuum gaps D25, D41 and D47\,au, with their emission dropping below detectable levels between 60--70\,au.  Figure \ref{fig:radial_profiles} (top) shows a comparison between our fiducial \texttt{frank} intensity profile and the \citet{Huang2018} observations calculated with \texttt{GoFish} \citep{gofish} assuming the geometry discussed above.  The profiles show excellent agreement both in terms of absolute flux level, slope, and substructure.  Even with the limited spatial resolution of our observations (a CLEAN beam size of $\sim$22\,au) we recover the D25 gap, however the D41 and D47 gaps are likely too shallow and narrow to be visible in our data.  We highlight the respective rms of our observations, which is a factor $\sim$100 lower, allowing us to detect the millimetre continuum emission beyond 60--70\,au.

We can also compare our millimetre continuum intensity profile with observations at other wavelengths, particularly those in scattered light.  \citet{Debes2017} present observations of TW Hya with HST/STIS in the R band (1.6\micron) tracing emission out to 180\,au, noting the presence of a wide gap at a radius of 1.43\arcsec\ (85\,au).  Similarly, \citet{vanBoekel2017} observed TW Hya with VLT/SPHERE at H band (1.6\micron) detecting a gap centred at 1.52\arcsec\ (90.5\,au).  Figure \ref{fig:radial_profiles} (bottom) shows the normalised radial profiles of these data scaled by $R^{2}$ to account for geometric dilution of the illuminating starlight.  The location of our newly-identified millimetre gap (D82) is coincident with the location of these scattered light gaps to within several au.  Variations between the precise gap locations are not unexpected, given that the millimetre continuum and scattered light observations are tracing different sized grains and also due to different physical mechanisms (thermal emission and scattered light, respectively).

\begin{figure*}
    \centering
    \includegraphics[width=\textwidth]{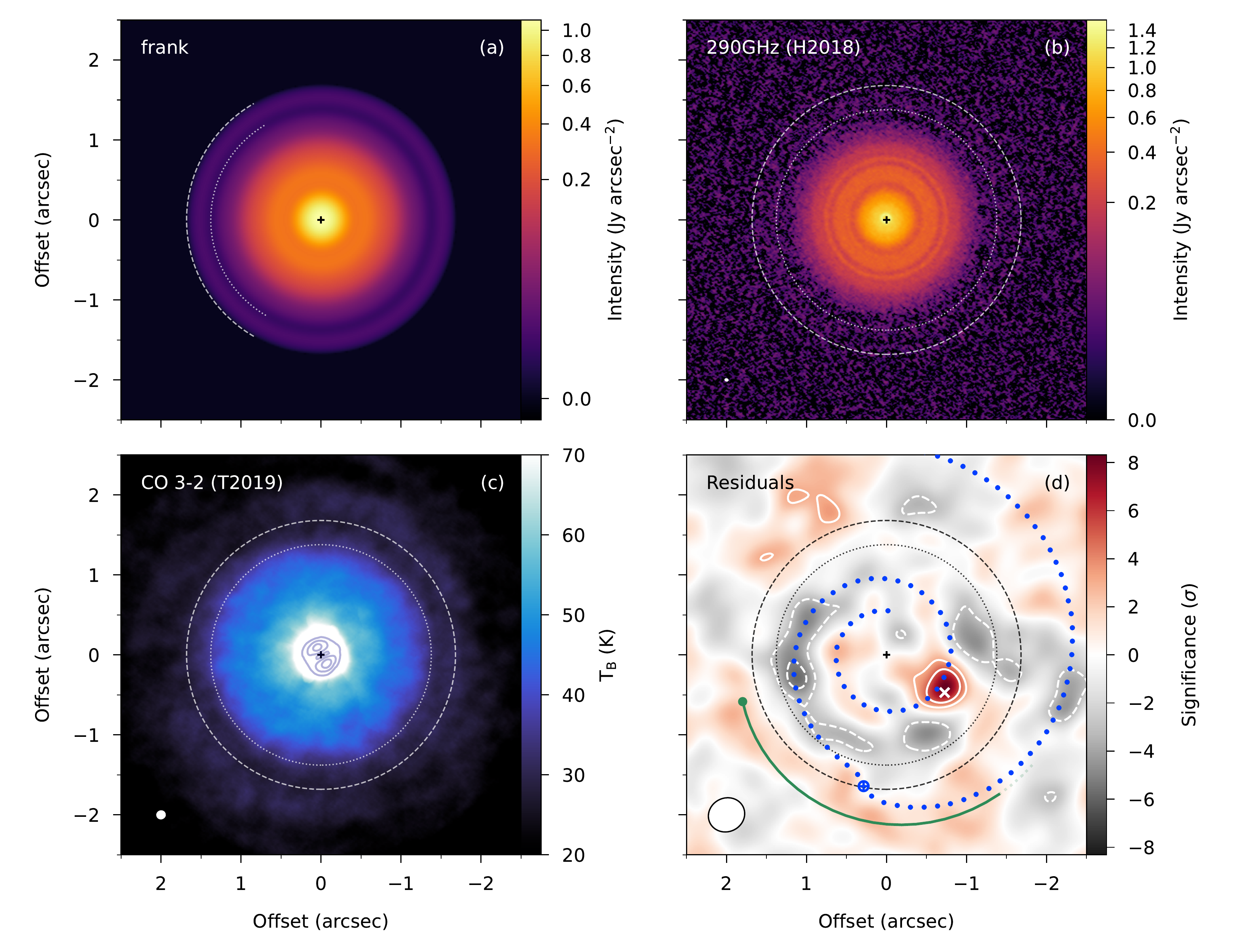}
    \caption{\textbf{(a)}: The fiducial \texttt{frank} intensity profile swept in azimuth and reprojected to the observed inclination and position angle of TW Hya.  Overlaid on all panels are ellipses showing the radial location of the gap at 82\,au (dotted) and 'edge' of the disc at 100\,au (dashed).
    \textbf{(b)}: 290\,GHz continuum image of TW Hya at a resolution of 0.13\arcsec \citep{Huang2018}.   
    \textbf{(c)}: Brightness temperature of the $^{12}$CO $J=3$--2 emission toward TW Hya \citep{Teague2019}, clipped between 20--70\,K to highlight emission in the outer disc.  The gap at 82\,au coincides with a `break' in $^{12}$CO emission.
    {\textbf{(d)}: Imaged residuals from our fiducial \texttt{frank} fit, where contours of $\pm$3, $5\sigma$ are shown ($\sigma=48$\,$\mu$Jy~beam$^{-1}$). A white cross marks the position of the localised continuum excess identified by \citet{Tsukagoshi2019} and a green arc traces the position of spiral structure identified in $^{12}$CO emission by \citet{Teague2019}.  Finally, a blue dotted line shows the shape of a spiral launched by a hypothetical planet (location marked with $\oplus$) as modelled by \citet{Sturm2020}.}}
    \label{fig:context}
\end{figure*}

We can also examine how our derived intensity profile would appear on the sky.  Figure \ref{fig:context}a shows the profile swept in azimuth and reprojected to the observed position angle and inclination of TW Hya.  Figure \ref{fig:context}b shows the corresponding 290\,GHz continuum image from \citet{Huang2018} overlaid with the location of the D82 gap and profile edge at 100\,au, which lie beyond the radial extent probed by these high resolution observations.

Given the extended radial range across which our newly identified substructure lies, we can compare it to observations of molecular line emission at similarly large radii.  Figure \ref{fig:context}c shows the brightness temperature of the $^{12}$CO $J=3$--2 emission toward TW Hya as presented in \citet{Huang2018} and \citet{Teague2019}.  The intensity scale has been cropped to highlight the `break' in emission between 80--100\,au, the beginning of which is coincident with the continuum gap D82.  It is also interesting to note that the D82 gap shares a similar radial location to the gap in CS $J=5$--4 emission at $\sim$1.6\arcsec\ (95\,au) presented in \citet{Teague2017}.  In addition, \citet{Nomura2021} recently presented high resolution maps of $^{13}$CO $J=3$--2, C$^{18}$O $J=3$--2, CS $J=7$--6 and CN $N=3$--2 toward TW Hya, all of which show breaks in emission between 80--90\,au that are coincident with D82.  Taken together, these observations are consistent with a scenario in which there is a large gas surface density depletion close to these radii, and that slight offsets can be attributed to imaging limitations or the fact that each molecular line is probing a slightly different region of the depletion.  Therefore, it appears that our millimetre continuum intensity profile shares a similar structure to both the gas and micron-sized dust in the outer disc.

\subsection{The extent of the millimetre continuum emission}

Our detection of millimetre emission in the outer disc has several implications.  Firstly, our observations appear to support the hypothesis of \citet{Rosotti2019_radii}, in that we confirm that deep observations reveal faint continuum emitting material beyond what has been previously reported as the outer edge of millimetre dust in the TW Hya disc.  However, our observations are unable to differentiate whether this emission originates from grains of a different size that would support the idea of an `opacity cliff', or whether the faint millimetre emission reflects a change in surface density.  To investigate this, similarly deep observations at multiple wavelengths would be required to calculate a spectral index in these outer disc regions.  

Secondly, disc sizes are commonly calculated with respect to some radius enclosing a percentage of total flux, e.g. the 68 per cent radius ($R_{68}$). { Our \texttt{frank} radial intensity profile has $R_{68} = 44.2$\,au, very close to previous measurements for TW Hya at similar frequencies \citep[e.g.\ $45.8\pm0.6$\,au;][]{Tripathi2017}.  However, this is clearly not an accurate representation of the `true' continuum disc based on our results.  In order to capture the larger extent enclosing this faint outer emission, much higher percentage flux radii are required (e.g. we find $R_{95} = 64.8$\,au, and $R_{99} = 88.1$\,au). } If this faint millimetre emission at large radii is a common feature amongst many other protoplanetary discs beyond TW Hya, then this would effectively steepen correlations between disc size and other quantities (such as mass and luminosity) based on observations with lower sensitivity \citep[e.g.][]{Andrews2018size,Hendler2020}.

\subsection{Characteristics of the outermost gap at 82 au}

The potential detection of the millimetre counterpart of the outermost gap in TW Hya raises questions as to its origin, in particular whether such a gap could be opened by a planet in the disc.  Following the procedure of \citet{Rosotti2016}, we measure a gap width of $11.7\pm3.2$\,au at 82\,au, which translates to a planet mass of $10.4\pm4.8$\,M$_{\oplus}$.  Previous studies have performed a similar analysis on the gap observed in scattered light.  \citet{Debes2013} placed an upper limit of the mass of the planetary companion of between 6--28\,M$_{\oplus}$ based on \emph{HST} observations, and \citet{vanBoekel2017} found their observed VLT/SPHERE gap depth could be reproduced by a 34\,M$_{\oplus}$ mass planet.  Taken as a range of possibilities, these values are in agreement with our derived planet mass based on the millimetre continuum.  Taking a slightly different approach, \citet{Mentiplay2019} used hydrodynamic models coupled with radiative transfer to simulate multiple observations of TW Hya.  They found that planets with masses between 32--95\,M$_{\oplus}$ were able to reproduce the outermost gap seen in scattered light, but note that such planets excite a spiral arm in the disc that should be apparent in scattered light observations (but is as yet undetected).  These masses are somewhat higher than our estimate from D82, but are within a factor of $\sim$3.  The addition of the extra constraints provided by our intensity profile between 60--100\,au may help refine this planet mass for future studies.      

\subsection{Investigating non-axisymmetric structure}

Despite the axisymmetric nature of our intensity profile, we can investigate any potential non-axisymmetric features in the observations by imaging the residual visibilities of the fiducial \texttt{frank} fit.  We adopted the same imaging parameters as outlined in Section \ref{sec:obs}, but explored various robust parameters, finding that $R=2$ provided the best image fidelity (with beam size $0.46\times0.41$ arcsec and position angle $-71.0\degr$).  The resulting residual image is shown in Figure \ref{fig:context}d.  The strongest residual feature is a localised excess at a position of $\Delta {\rm RA} = -0.74\arcsec$, $\Delta \rm{Dec} = 0.38\arcsec$, corresponding to a radial distance of $50\pm4$\,au ($\mathrm{P.A.}=242.5\degr$) from the central star.  It is detected at approximately $8\sigma$ with an integrated flux density of $0.46\pm0.16$\,mJy.  A blue cross marks the position of the small scale continuum excess first identified in \citet{Tsukagoshi2019}, coincident with the peak of our excess within 0.05\arcsec (3\,au, close to the astrometric precision of ALMA).  Estimating the dust mass of the continuum excess assuming optically thin emission and adopting the same assumptions as \citet{Tsukagoshi2019} (but extrapolated to our observing frequency) gives a value of $(3.1\pm1.1) \times 10^{-2}$\,M$_{\oplus}$, in good agreement with their dust mass of $(2.83\pm0.06)\times10^{-2}$\,M$_{\oplus}$.  This suggests our deep observations are tracing the same feature within the disc, which is impressive given the size of our nominal beam (0.44\arcsec, 26\,au) compared to the resolved size of the excess (2--3\,au).

In addition to the localised excess, there are several arc-like structures in the residuals between radii of 40--120\,au detected with a S/N of between 2--6.  We can compare their morphology to non-axisymmetric structure that has been previously identified in the TW~Hya disc.  \citet{Teague2019} identified spiral structure in the intensity of $^{12}$CO $J=3$--2 emission in the disc, overlaid Figure\ \ref{fig:context}d as a green line, which appears coincident with our continuum residuals in this region.  \citet{Sturm2020} further analysed these residuals, finding that a hydrodynamic model with planet at a radius of 1.65\arcsec\ (100\,au) would generate a spiral wake consistent with their observed morphology.  We have overlaid this spiral morphology on Figure\ \ref{fig:context}d with a blue dotted line, demonstrating remarkable agreement with the continuum residual arcs.  It therefore appears that our residual map is detecting the millimetre continuum counterpart to the spiral structure previously identified in $^{12}$CO molecular line emission.

\section{Conclusions}

We have analysed deep millimetre observations of TW~Hya using the visibility modelling code \texttt{frank}. We summarise our findings below:  

\begin{itemize}
    
    \item We detect faint, but structured, millimetre continuum emission out to radius of approximately 100\,au.  While our 68 per cent flux radius remains relatively unchanged with respect to previous estimates, our 95 and 98 per cent flux radii are 64.8 and 88.1\,au, respectively, suggesting that these metrics may be more appropriate measures of the true sizes of discs at millimetre wavelengths. 
        
    \item Our findings support the hypothesis of \citet{Rosotti2019_radii} in that previous millimetre continuum observations of discs are limited by sensitivity. If ubiquitous, this has implications for the many scaling relations derived from the continuum sizes of protoplanetary discs.
        
    \item Our derived intensity profile shows tentative evidence of a gap in the millimetre dust disc at 82\,au (D82) and a bright ring at 91\,au (B91), although their exact properties are dependent on the behaviour of the visibilities at unobserved baselines. The gap at 82\,au is coincident with several other features observed in both the gas and dust of the TW Hya disc, including a gap in scattered light and several molecular lines.  
    
    \item If a planet were to be responsible for opening such a gap, then it would be a sub-Neptune with mass $10.4\pm4.8$\,M$_{\oplus}$, in broad agreement with similar analyses of multiple scattered light observations, and forward hydrodynamic modelling.
    
    \item Examining our fit residuals, we confirm the detection of the localised au-scale millimetre continuum excess at 52\,au in the TW Hya disc first reported by \citet{Tsukagoshi2019} and derive a similar dust mass of $(3.1\pm1.1) \times 10^{-2}$\,M$_{\oplus}$. We also find tentative hints of non-axisymmetric spiral structure, which may be the continuum counterpart to similar structures observed in $^{12}$CO emission.
    
\end{itemize}

The nature of the radial extension of the millimetre continuum could be further probed by sensitive multi-frequency observations to shed light on the properties of the dust emitting in the outer disc.  In addition, complementary analysis techniques in both the visibility and image plane may help to further refine the properties of substructure in these regions.  Nevertheless, our results demonstrate the utility of deep millimetre observations at intermediate spatial resolution which, when combined with super-resolution analysis techniques, can \emph{i}) recover emission from the faint outer regions of protoplanetary discs, and \emph{ii}) characterise any disc substructure.  It is likely that this technique can be extended to study the outer regions of many other discs that exhibit substructure in scattered light observations \citep[e.g.][]{Avenhaus2018} beyond what would normally considered the edge of the millimetre dust disc.

\section*{Acknowledgements}

We are grateful to Jane Huang, Roy van Boekel and John Debes for making their data available. We thank the referee, Simon Casassus, for a thorough report.
JDI acknowledges support from the STFC under ST/R000287/1.  CW acknowledges support from the University of Leeds, STFC and UKRI (grant numbers ST/T000287/1, MR/T040726/1). RAB is supported by a Royal Society University Research Fellowship. GPR acknowledges support from the Netherlands Organisation for Scientific Research (NWO, program number 016.Veni.192.233) and from an STFC Ernest Rutherford Fellowship (grant number ST/T003855/1).
This paper makes use of the following ALMA data: ADS/JAO.ALMA\#2016.1.00464.S. 

\section*{Data Availability}

The data underlying this article will be shared on reasonable request.



\bibliographystyle{mnras}



\bsp	
\label{lastpage}
\end{document}